\documentclass[aps,prl,graphicx,twocolumn,showpacs,preprintnumbers,amsmath,amssymb,superscriptaddress]{revtex4-1}

\usepackage{graphicx}

\begin{document}

\title{Linear and quadratic magnetoresistance in the semimetal SiP$_{2}$}

\author{Yuxing Zhou}
\author{Zhefeng Lou}
\affiliation{Department of Physics, Zhejiang University, Hangzhou $310027$, China}
\author{ShengNan Zhang}
\affiliation{Institute of Physics,  \'{E}cole Polytechnique F\'{e}d\'{e}rale de Lausanne (EPFL), CH-1015 Lausanne, Switzerland}
\affiliation{National Centre for Computational Design and Discovery of Novel Materials MARVEL, \'{E}cole Polytechnique F\'{e}d\'{e}rale de Lausanne (EPFL), CH-1015 Lausanne, Switzerland}
\author{Huancheng Chen}
\author{Qin Chen}
\author{Binjie Xu}
\affiliation{Department of Physics, Zhejiang University, Hangzhou $310027$, China}
\author{Jianhua Du}
\affiliation{Department of Applied Physics, China Jiliang University, Hangzhou $310018$, China}
\author{Jinhu Yang}
\affiliation{Department of Physics, Hangzhou Normal University, Hangzhou $310036$, China}
\author{Hangdong Wang}
\affiliation{Department of Physics, Hangzhou Normal University, Hangzhou $310036$, China}
\author{Chuanying Xi}
\affiliation{Anhui Province Key Laboratory of Condensed Matter Physics at Extreme Conditions, High Magnetic Field Laboratory, Chinese Academy of Sciences, Hefei 230031, China}
\author{Li Pi}
\affiliation{Anhui Province Key Laboratory of Condensed Matter Physics at Extreme Conditions, High Magnetic Field Laboratory, Chinese Academy of Sciences, Hefei 230031, China}
\affiliation{Hefei National Laboratory for Physical Sciences at Microscale, University of Science and Technology of China, Hefei 230026, China}
\author{QuanSheng Wu}
\author{Oleg V. Yazyev}
\affiliation{Institute of Physics, \'{E}cole Polytechnique F\'{e}d\'{e}rale de Lausanne (EPFL), CH-1015 Lausanne, Switzerland}
\affiliation{National Centre for Computational Design and Discovery of Novel Materials MARVEL, \'{E}cole Polytechnique F\'{e}d\'{e}rale de Lausanne (EPFL), CH-1015 Lausanne, Switzerland}
\author{Minghu Fang}\email{Corresponding author: mhfang@zju.edu.cn}
\affiliation{Department of Physics, Zhejiang University, Hangzhou $310027$, China}
\affiliation{Collaborative Innovation Center of Advanced Microstructures, Nanjing University, Nanjing $210093$, China}

\date{\today}

\begin{abstract}
Multiple mechanisms for extremely large magnetoresistance (XMR) found in many topologically nontrivial/trivial semimetals have been theoretically proposed, but experimentally it is unclear which mechanism is responsible in a particular sample. In this article, by the combination of band structure calculations, numerical simulations of magnetoresistance (MR), Hall resistivity and de Haas-van Alphen (dHvA) oscillation measurements, we studied the MR anisotropy of SiP$_{2}$ which is verified to be a topologically trivial, incomplete compensation semimetal. It was found that as magnetic field, \emph{H}, is applied along the \emph{a} axis, the MR exhibits an unsaturated nearly linear \emph{H} dependence, which was argued to arise from incomplete carriers compensation. For the \emph{H} $\parallel$ [101] orientation, an unsaturated nearly quadratic \emph{H} dependence of MR up to 5.88 $\times$ 10$^{4}$$\%$ (at 1.8 K, 31.2 T) and field-induced up-turn behavior in resistivity were observed, which was suggested due to the existence of hole open orbits extending along the $k_{x}$ direction. Good agreement of the experimental results with the simulations based on the calculated Fermi surface (FS) indicates that the topology of FS plays an important role in its MR.
\end{abstract}

\pacs{}

\maketitle

\section{\romannumeral1. INTRODUCTION}
Since magnetoresistance (MR) has a great potential in applications such as hard drives \cite{daughton1999gmr} and magnetic sensors \cite{reig2009magnetic}, the search for new materials with large MR has attracted much attention in the past decades. Though the well-known giant magnetoresistance (GMR) in magnetic multilayers \cite{PhysRevLett.61.2472,PhysRevB.39.4828} and the colossal magnetoresistance (CMR) in perovskite manganites \cite{RevModPhys.73.583} have been widely exploited, recent discoveries of the materials with extremely large magnetoresistance (XMR) up to 10$^{6}\%$ rekindled the enthusiasm for MR research. XMR has been observed in elements and compounds, such as Bi \cite{PhysRev.101.544}, graphite \cite{PhysRevB.25.5478}, $\alpha$-Ga \cite{chen2018large}, Dirac semimetal Na$_{3}$Bi \cite{PhysRevB.85.195320,xiong2015evidence}, and Cd$_{3}$As$_{2}$ \cite{PhysRevLett.113.246402,liang2015ultrahigh}, Weyl semimetals of TaAs family \cite{shekhar2015extremely,PhysRevX.5.031023,PhysRevB.92.205134,du2016large}, WTe$_{2}$ \cite{ali2014large}, and $\beta$-WP$_{2}$ \cite{PhysRevB.96.121107,PhysRevLett.117.066402,PhysRevB.96.121108}, transition metal dipnictides such as TPn$_{2}$ (T = Ta and Nb, Pn = P, As and Sb) \cite{PhysRevB.93.195119,doi:10.1063/1.4940924,PhysRevB.93.195106,PhysRevB.94.041103,PhysRevB.94.121115,PhysRevB.93.184405,TaAs2}, $\alpha$-WP$_{2}$ \cite{PhysRevB.97.245101}, rock salt rare earth compound LaBi/Sb \cite{tafti2016resistivity,PhysRevB.93.241106,PhysRevLett.117.127204} and others.

Several mechanisms have been proposed to explain the XMR found in these semimetals including topologically nontrivial or trivial materials. Nontrivial band topology inducing linear band dispersion is believed to be responsible for the linear field dependent MR such as in Cd$_{3}$As$_{2}$ \cite{PhysRevLett.113.246402,liang2015ultrahigh}. The classical carrier compensation scenario can be used to explain the non-saturating quadratic dependence of MR such as in WTe$_{2}$ \cite{ali2014large}. An angle-resolved photoemission spectrometry (ARPES) measurement on WTe$_{2}$ \cite{PhysRevLett.113.216601} confirmed that the temperature dependent band structure is consistent with the MR measurements, thus giving evidence to support the carrier compensation theory. However, although LaAs shows nearly perfect carrier compensation, the magnitude of MR is much smaller, which is believed to be caused by the electron and hole mobility mismatch \cite{PhysRevB.96.235128}. Recent ARPES results on MoTe$_{2}$, which has a crystal structure identical to that of WTe$_{2}$, illustrated that the net size of hole pockets is larger than the net size of electron pockets, indicating the compensation mechanism is invalid for the non-saturating XMR of MoTe$_{2}$ \cite{PhysRevB.95.241105}. YSb lacks topological protection and perfect electron-hole compensation, but also exhibits XMR behavior \cite{PhysRevLett.117.267201}. A small difference between the concentrations of electrons and holes will lead to saturation of MR at high magnetic field such as in Bi \cite{PhysRev.101.544} and graphite \cite{PhysRevB.25.5478}. Zhang \emph{et al}. \cite{PhysRevB.99.035142} showed that MR has a quadratic relation in weak magnetic field, but saturates in high field if the FS is closed and the saturation value is determined by the difference in charge carrier concentrations. The other mechanism attributes the XMR to open-orbit trajectories of charge carriers driven by Lorentz force under magnetic field as a result of non-closed Fermi surface (FS) \cite{chambers2012electrons,pippard1989magnetoresistance,PhysRevLett.111.056601}. Experimentally, it is difficult to identify which mechanism is responsible for MR in a particular sample. It is necessary to make a clear connection between experimental observations and theoretical models.

SiP$_{2}$ crystallizes in a cubic pyrite-type structure \cite{donohue1968preparation} with space group \emph{P}a$\bar{3}$, and was recently discovered to be a promising negative electrode material for Li- and Na-ion batteries due to its outstanding capacity \cite{duveau2016pioneer}. In contrast to the isostructural NiP$_{2}$, PtP$_{2}$, or pyrite itself (FeS$_{2}$) being semiconductors, SiP$_{2}$ is characterized as a semimetal with nearly filled Brillouin zone \cite{donohue1968preparation}. To understand its semimetal character, the electronic structure of SiP$_{2}$ had been calculated \cite{bachhuber2011first,farberovich1979fermi,farberovich1976density,farberovich1975band}. It was suggested by Bachhuber \emph{et al}. \cite{bachhuber2011first} that a flat band segment occurs between the $\Gamma$ and X point, resulting in no gap formed.

In this article, we have successfully grown high-quality SiP$_{2}$ crystals, measured their longitudinal resistivity with various magnetic field orientations, Hall resistivity, de Haas-van Alphen (dHvA) oscillations occurring on isothermal magnetization and calculated its band structure. The results show SiP$_{2}$ is a topologically trivial and incomplete compensation semimetal. It was found that the MR exhibits remarkable anisotropy. As \emph{H} is applied in the \emph{a} axis, a non-saturating linear field dependence of MR with relatively small value (5.96 $\times$ 10$^{2}\%$ at 2 K, 9 T) occurs. While \emph{H} is applied in the [101] direction, MR (2.17 $\times$ 10$^{3}\%$ at 2 K, 9 T) exhibits a non-saturating quadratic \emph{H} dependence. The mechanisms of the two types MR will be discussed.

\section{\romannumeral2. EXPERIMENTAL METHODS AND CALCULATIONS}

Single crystals of SiP$_{2}$ were grown by a chemical vapor transport method. High purity Si and P powder were mixed in the mole ratio 1 : 2, then sealed in an evacuated silica tube with PBr$_{5}$ producing enough Br$_{2}$ as a transport agent. The quartz tube was placed in a tube furnace with a temperature gradient of 1200 $^{\circ}$C to 800 $^{\circ}$C for one week. The black shiny SiP$_{2}$ crystals were obtained at the cool end of the silica tube. A single crystal with dimensions of 1 $\times$ 1 $\times$ 0.15 mm$^{3}$ and crystalline cleavage surface (200) [see in Fig. 1(b)] was selected for transport and magnetic measurements. The crystal structure was determined by X-ray diffraction (XRD) measurements using a PANalytical diffractometer. The powder XRD pattern is shown in Fig. 1(c), which confirms that SiP$_{2}$ crystallizes in a pyrite-type structure. The fit to XRD data yields the lattice parameters: \textit{a} = \textit{b} = \textit{c} = 5.704(9) $\rm {\AA}$ (weight profile factor $R_{wp}$ = 9.96$\%$ and the goodness-of-fit $\chi^{2}$ = 0.9229). A standard four-probe method was used for electrical resistivity measurements on a Physical Property Measurement System (Quantum Design, PPMS-9 T) and a water-cooled magnet with the highest magnetic field of 31.2 T. The magnetization measurements were performed on a Magnetic Property Measurement System (Quantum Design, MPMS-7 T).

\begin{figure}
  \centering
  \includegraphics[width=8cm]{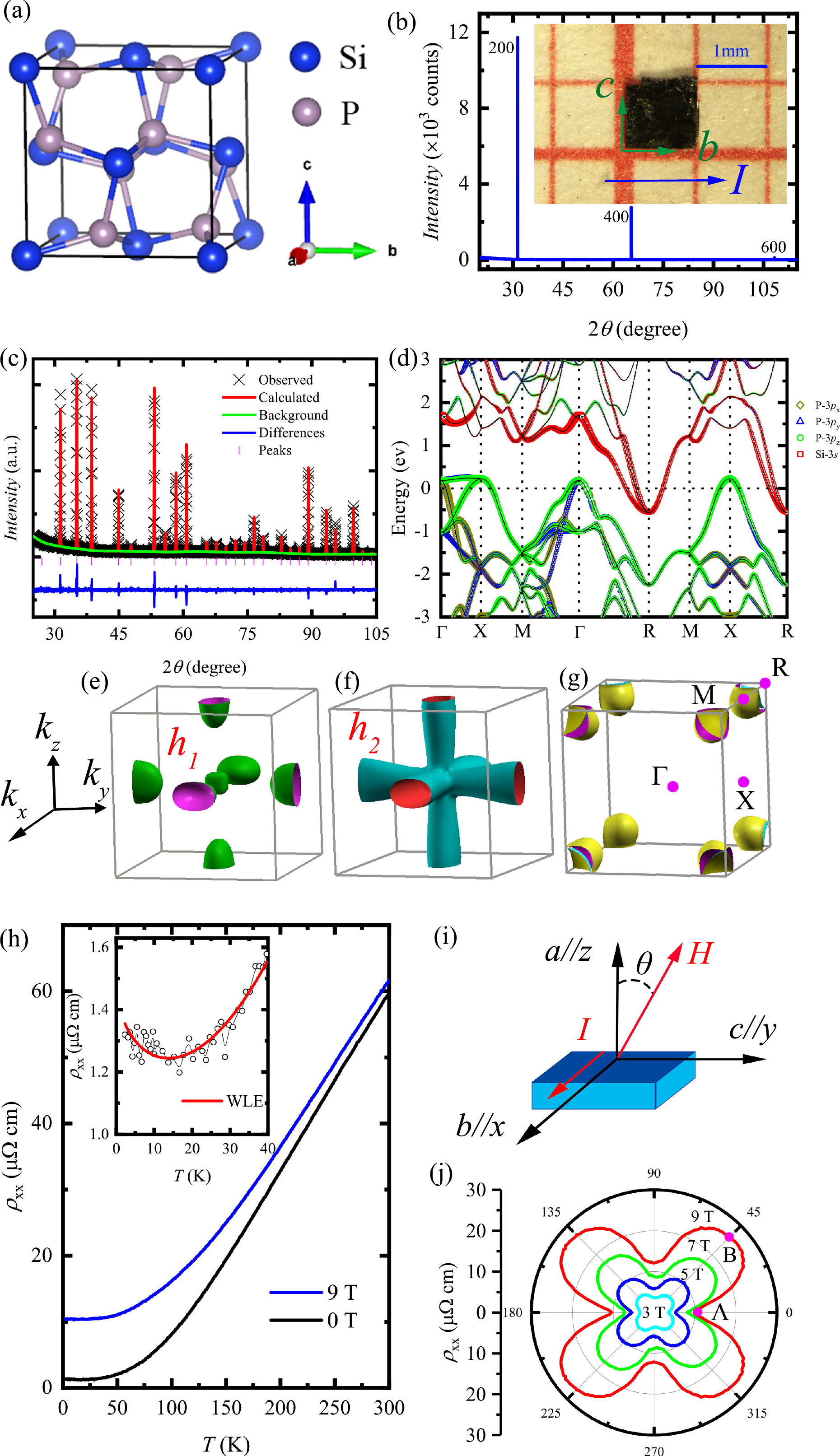}\\
  \caption{(a) Crystal structure of SiP$_{2}$. (b) XRD pattern and the photograph (inset) of a SiP$_{2}$ crystal. (c) XRD pattern of powder obtained by grinding SiP$_{2}$ single crystals. Its Rietveld refinement is shown by the red solid line. (d) Calculated band structure of SiP$_{2}$ without spin-orbit coupling (SOC) (no large difference with SOC due to its light elements, not shown). (e) and (f) 3D view of hole-type FSs and (g) electron-type FSs. (h) Temperature dependence of resistivity $\rho(T)$ of a SiP$_{2}$ crystal measured at 0 T and 9 T. The inset is the enlarged low temperature $\rho(T)$ data at 0 T, and the red line is WLE fitting. (i) Schematic diagram of MR measurements, the current is applied along the \textit{b} axis and the field angle, $\theta$, is given in the $a-c$ plane. (j) The angular polar plot of resistivity at 2 K measured under various fields.}\label{1}
\end{figure}

Meanwhile, we performed numerical simulations based on the Boltzmann transport theory and first-principles calculations \cite{PhysRevB.99.035142} that can be compared with the results of experimental measurements. The band structure is calculated using the generalized gradient approximation \cite{perdew1996generalized} within the VASP package \cite{PhysRevB.54.11169,PhysRevB.59.1758}. The Fermi surface and transport calculation are performed with WannierTools \cite{mostofi2014updated} package which is based on the maximally localized Wannier function tight-binding model \cite{PhysRevB.56.12847,PhysRevB.65.035109,marzari2012maximally} constructed by using the Wannier90 \cite{wu2018wanniertools} package.

Within the relaxation time approximation, the band-wise conductivity tensor $\bold{\sigma}$ is calculated by solving the Boltzmann equation in presence of an applied magnetic field as \cite{ashcroft1976solid,PhysRevB.99.035142,PhysRevB.79.245123},

\begin{equation}
\sigma^{(n)}_{ij}(\bold{B})=\frac{ e^2}{4 \pi^3} \int d\bold{k} \tau_n \bold{v}_n(\bold{k})  \bold{\bar{v}}_n(\bold{k})
\left(- \frac{\partial f}{\partial \varepsilon} \right)_{\varepsilon=\varepsilon_n(\bold{k})},
\label{eqn-sigmaij}
\end{equation}

where $e$ is the electron charge, $n$ is the band index, $\tau_n$ is the relaxation time of $n$th band that is assumed to be independent on the wavevector $\bold{k}$, $f$ is the Fermi-Dirac distribution, $\bold{v}_n(\bold{k})$ is the velocity defined by the gradient of band energy,
\begin{equation}
\bold{v}_n(\bold{k})=\frac{1}{\hbar} \nabla_{\bold{k}} \varepsilon_n(\bold{k}),
\label{eqn-velocity}
\end{equation}
and $\bar{\bold{v}}_n(\bold{k})$ is the weighted average of velocity over the past history of the charge carrier,
\begin{equation}
\bar{\bold{v}}_n(\bold{k}) = \int^0_{-\infty} \frac{dt}{\tau_n} e^{\frac{t}{\tau_n}} \bold{v}_n(\bold{k}_{n}(t)) .
\label{eqn-aver_velo}
\end{equation}
The orbital motion of charge carriers in applied magnetic field causes the time evolution of $\bold{k}_n(t)$, written as,
\begin{equation}
\frac{d \bold{k}_n(t)}{dt} = - \frac{e}{\hbar} \bold{v}_n(\bold{k}_{n}(t)) \times \bold{B}
\label{eqn-evol_k}
\end{equation}
with $ \bold{k}_n(0)=\bold{k}$.
The total conductivity is the sum of band-wise conductivities, i.e. $\sigma_{ij} = \sum_n \sigma^{(n)}_{ij} $, which is then inverted to obtain the resistivity tensor $\hat{\rho} = \hat{\sigma}^{-1}$.

\section{\romannumeral3. RESULTS AND DISCUSSIONS}

In order to explore the role of the Fermi surface topology in MR, we calculated the band structure and FS of SiP$_{2}$, as shown in Fig. 1(d) - (g). The FS is composed of two hole pockets near the $\Gamma$ point of P-3\emph{p} character and four electron pockets located at the R point of Si-3\emph{s} character, exhibiting three dimensional (3D) nature. The existence of both hole and electron pockets of the FS is consistent with SiP$_{2}$ being a semimetal and the calculation results reported by Bachhuber \emph{et al}. \cite{bachhuber2011first}. In addition, it should be pointed out that no crossing between the conduction and valence bands emerges in the calculated band structure and all FS sheets have zero Chern number, indicating that SiP$_{2}$ is a topologically trivial semimetal.

\begin{figure}
  \centering
  \includegraphics[width=8cm]{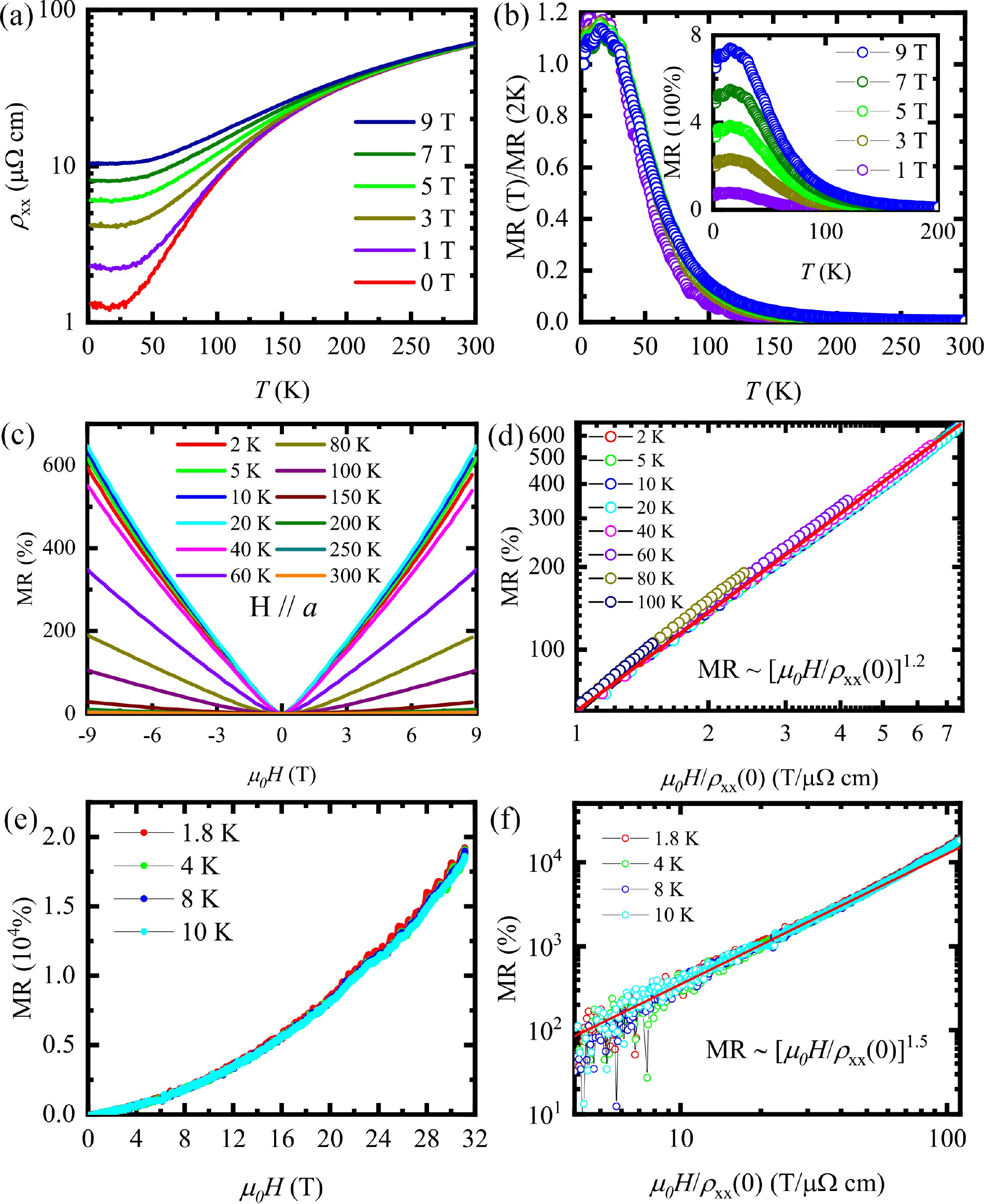}\\
  \caption{(a) Temperature dependence of resistivity measured at various magnetic fields applied along the \emph{a} axis. (b) The normalized MR vs. temperature under various magnetic fields. The inset is MR as a function of temperature. (c) MR of SiP$_{2}$ measured under different temperatures with the field along the \textit{a} axis. (d) Kohler scaling analysis on the MR data measured on PPMS, the solid red line indicates the fitting of Kohler's rule with \emph{m} = 1.2. (e) Field dependence of MR of SiP$_{2}$ measured near $\theta = 0^{\circ}$ ($\pm$ 7 $^{\circ}$, see the text)at different temperatures up to 31.2 T. (f) Kohler scaling analysis on the MR data measured on a water cooled magnet, the solid red line indicates the fitting of Kohler's rule with \emph{m} = 1.5.}\label{2}
\end{figure}
\begin{figure}
  \centering
  \includegraphics[width=8cm]{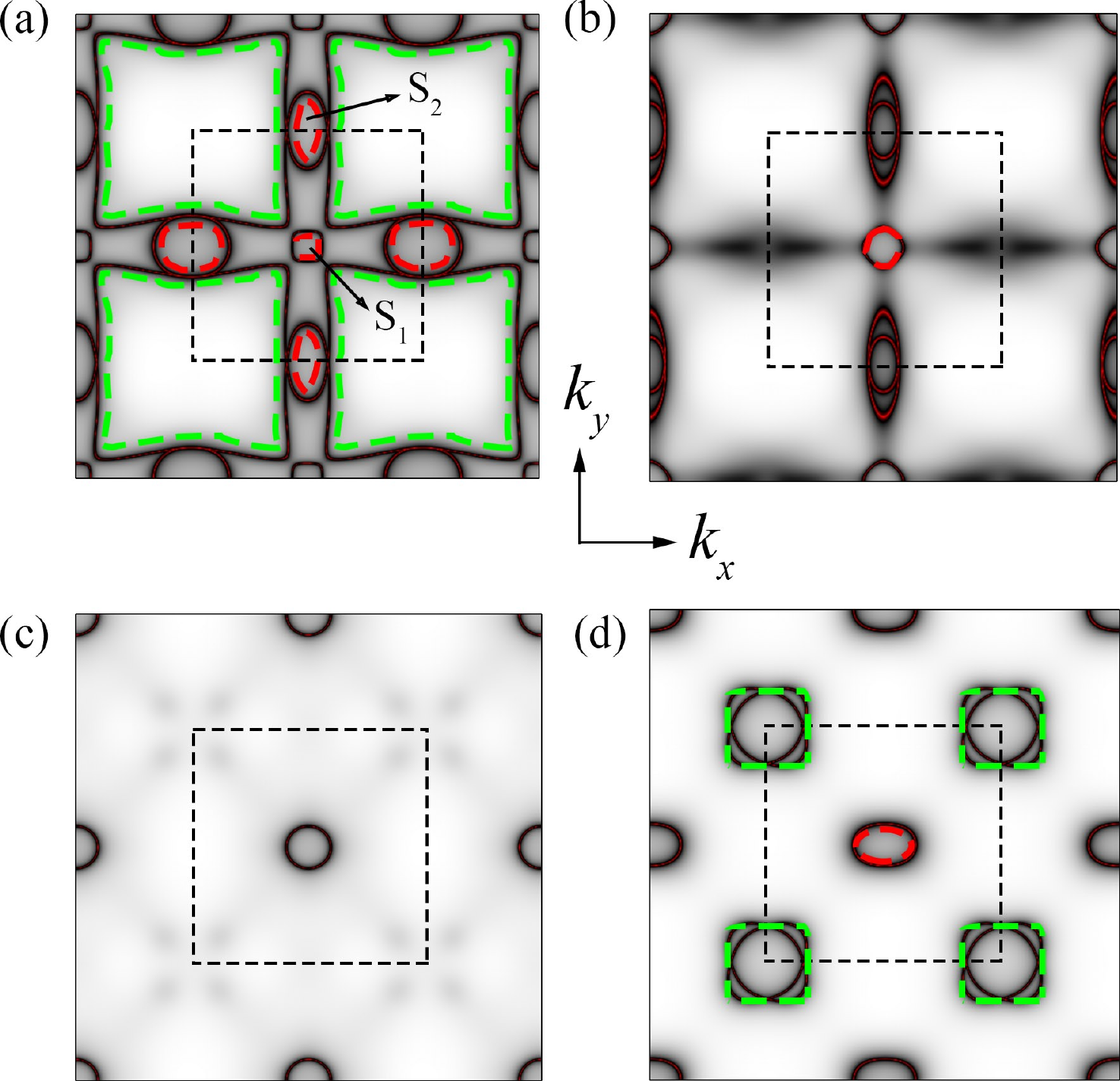}\\
  \caption{Typical cross-sections of the FS of SiP$_{2}$ in $k_{x}-k_{y}$ plane corresponding to (a) $k_{z}$ = 0, (b) $k_{z}$ = 0.2$\pi/a$, (c) $k_{z}$ = 0.5$\pi/a$, (d) $k_{z}$ = $\pi/a$. Red and green dashed lines highlight the closed hole and electron orbits, respectively. The black dashed squares indicate the first Brillouin zone.}\label{3}
\end{figure}

Figure 1(h) shows the temperature dependence of resistivity, $\rho(T)$, measured with current \emph{I} along the \emph{b} axis and at both magnetic field $\mu_{0}H$ = 0 T and 9 T applied along the \emph{a} axis, respectively. At $\mu_{0}H$ = 0 T, the resistivity decreases monotonously with decreasing temperature above 15 K, and reaches a minimum at 15 K [see Fig. 1(h), inset], then increases slightly at low temperature. We suggest that the emergence of minimum at \emph{T} = 15 K in $\rho(T)$ may result from the well-known weak localization effect (WLE) \cite{RevModPhys.57.287,altshuler1985electron,Abrikosov1990Fundamentals}, which arises from the carriers backscattered coherently by randomly distributed disorder existing in the crystals, and had been used to explain a similar behavior in some oxides, such as SrRuO$_{3}$ \cite{PhysRevB.67.174423} and LaNiO$_{3}$ \cite{herranz2004weak} thin films.  As discussed by Herranz \emph{et al}. \cite{PhysRevB.67.174423, herranz2004weak}, we fitted the $\rho(T)$ data at lower temperatures by using the equation \cite{PhysRevB.67.174423,herranz2004weak}:

\begin{eqnarray}
  \rho = \frac{1}{\sigma_{0}+aT^{1/2}} + bT^{2}
\end{eqnarray}

The first term is related to quantum corrections to the conductivity in 3D systems, the second term in $T^{2}$ is included to extend analytical description to higher temperatures. As shown in the inset of Fig. 1(h), Eq. (1) can well describe the $\rho(T)$ data at low temperatures with the fitting parameters $\sigma_{0}$ = 6.1$\times 10^{5}$ $\Omega^{-1}$ cm$^{-1}$, \textit{a} = 7.26$\times 10^{4}$ $\Omega^{-1}$ cm$^{-1}$ K$^{-1/2}$ and \textit{b} = 4.37$\times 10^{-10}$ $\Omega$ cm K$^{-2}$. We note that no peak in MR at \emph{H} = 0 T emerges in our SiP$_{2}$ crystals, which appears in some thin film samples with WLE \cite{PhysRevLett.97.016801,PhysRevB.25.2937,PhysRevB.86.075102}. The WLE results in a relatively low residual resistivity ratio (RRR) = $\rho(300K)/\rho(2K) \approx$ 45. At $\mu_{0}H$ = 9 T, $\rho(T)$ exhibit a metallic behavior in the whole temperature range (2 - 300 K), and the resistivity is remarkably enhanced, even at 300 K, implying that large MR occurs in this non-magnetic semimetal.

\begin{figure}
  \centering
  \includegraphics[width=8cm]{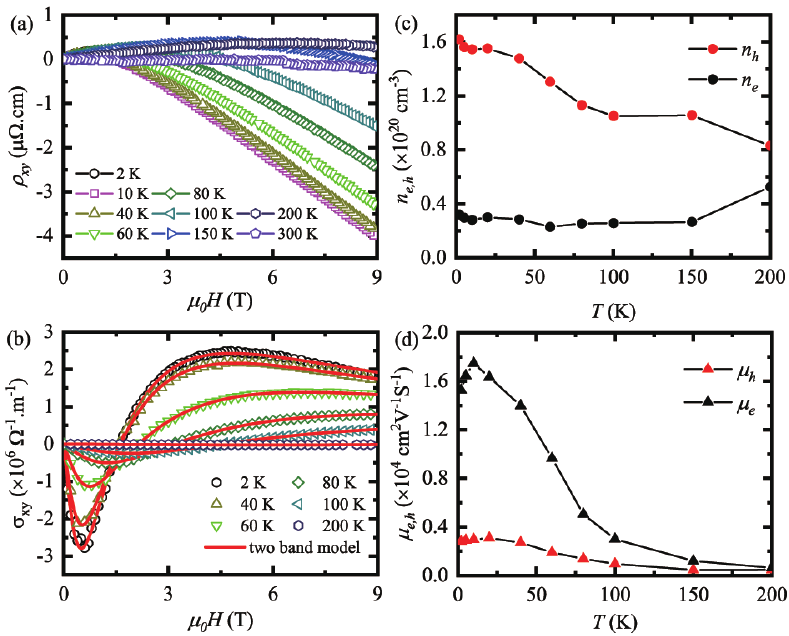}\\
  \caption{(a) Field dependence of Hall resistivity $\rho_{xy}$ measured at various temperatures (\emph{H} $\parallel$ \emph{a} axis). (b) Several $\sigma_{xy}(H)$ data at various temperatures with the fitting lines by using the two-band model (see text). (c) and (d) Temperature dependence of carrier concentrations and mobilities, respectively.}\label{4}
\end{figure}

Then, we measured the resistivity anisotropy at 2 K in $\mu_{0}H$ = 3, 5, 7 and 9 T with \emph{I} along the \emph{b} axis, and by rotating the magnetic field \emph{H} in the $a-c$ plane [see Fig. 1(i)]. Figure 1(j) shows the angular resistivity polar plot, which exhibits a fourfold symmetry, i.e. $\rho(\theta) = \rho(\theta+\pi/2)$, the resistivity grows quickly from a minimum at $\theta$ = 0$^{\circ}$ (\emph{H} $\parallel$ \emph{a} axis) to a maximum at $\theta$ = 45$^{\circ}$ [\emph{H} $\perp$ (101) plane], and then decreases rapidly to another minimum at $\theta$ = 90$^{\circ}$ (\emph{H} $\parallel$ \emph{c} axis), which is consistent with the cubic structure of SiP$_{2}$ crystal. As we know, the resistivity anisotropy reflects the symmetry of the FS projected onto the plane perpendicular to current. Compared with Cu crystal, a representative material \cite{PhysRevB.99.035142}, also crystallizing in cubic structure, SiP$_{2}$ has a simpler FS, and provides a clearer platform for studying MR mechanism based on FS topology. In order to reveal the physics underlying the MR anisotropy, we measured both the field and temperature dependencies of resistivity for the magnetic field orientations corresponding to extrema points marked by A and B in Fig. 1(j).

As \emph{H} is applied along the \emph{a} axis [$\theta = 0^{\circ}$, the A point in Fig. 1(j)] with a minimum resistivity relative to other orientations, the $\rho(T)$ measured at various fields is shown in Fig. 2(a). Although the resistivity is remarkably enhanced by magnetic field at lower temperatures, the field-induced up-turn was not observed, which is a typical behavior for many trivial or nontrivial semimetals with XMR \cite{PhysRevB.92.180402,PhysRevB.96.075132,PhysRevB.97.245101}. The normalized MR, with the conventional definition MR = $\frac{\Delta\rho}{\rho(0)} = [\frac{\rho(\emph{H}) - \rho(0)}{\rho(0)}] \times 100\%$, has the same temperature dependence at different magnetic fields [see Fig. 2(b)], excluding the existence of a magnetic field-dependent gap. Figure 2(c) shows the MR as a function of field at various temperatures. The MR reaches 5.96 $\times$ 10$^{2}\%$ at 2 K and 9 T, and does not show any sign of saturation up to the highest field in PPMS. The MR can be described by the Kohler scaling law \cite{PhysRevB.92.180402,ziman2001electrons}:

\begin{eqnarray}
  MR = \frac{\Delta\rho_{xx}(\emph{T},\emph{H})}{\rho_{0}(\emph{T})} = \alpha[\frac{\mu_{0}H}{\rho_{xx}(0)}]^{m}
\end{eqnarray}

As shown in Fig. 2(d), all MR data from 2 to 100 K collapse onto a single line plotted as MR $\sim$ $H/\rho(0)$ curve, with $\alpha$ = 56.4 ($\mu\Omega$ cm/T)$^{1.2}$ and \textit{m} = 1.2 obtained by fitting, indicating that MR has a nearly linear field dependence. To understand this nearly linear magnetic field dependence, we plot the representative orbits in Fig. 3(a) - (d). The circular orbits in Fig. 3(a) and the orbits in Fig. 3(b) - (d) can be attributed to closed electron (in green) and hole (in red) orbits. But the square like orbits [indicated by the green dashed line in Fig. 3(a)] are more complex, since they originate from joining the hole pocket fragments in the adjacent periodic replicas of the Brillouin zone. However, these square like orbits are electron orbits rather than hole orbits, since they enclose filled states. Therefore, the perfect compensation between the electron and hole charge carriers is altered upon applying magnetic field oriented along the $a$ axis ($\theta = 0^{\circ}$). The incomplete compensation induces the departure of resistivity from the ideal parabolic to nearly linear scaling.

\begin{figure}
  \centering
  \includegraphics[width=8cm]{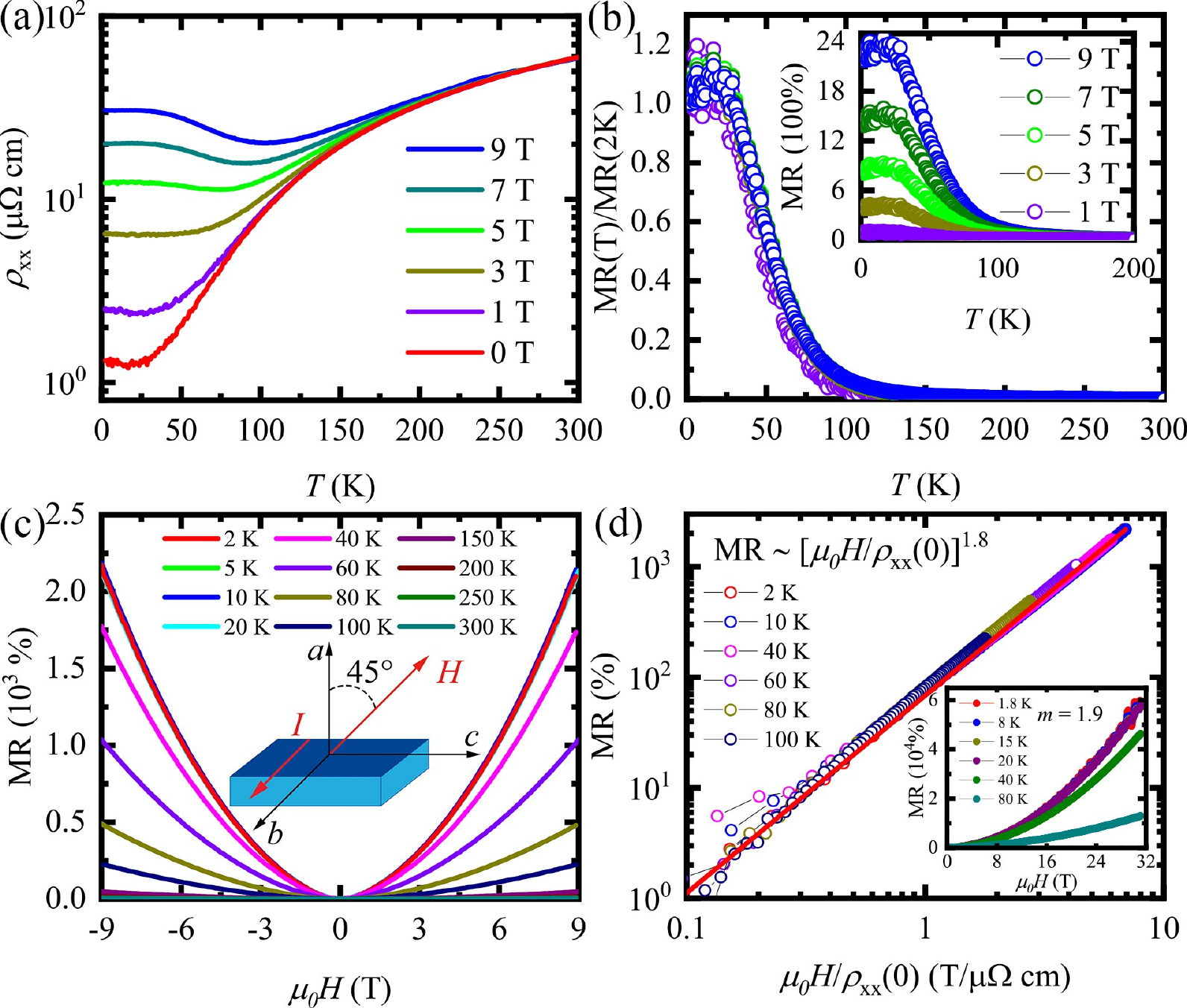}\\
  \caption{(a) Temperature dependence of resistivity measured with \emph{H} $\parallel$ [101]. (b) Normalized MR versus temperature at various magnetic fields. The inset is MR as a function of temperature. (c) Field dependence of MR of SiP$_{2}$ measured at different temperatures. The inset illustrates the direction of \emph{H} and \emph{I}. (d) Kohler scaling plots for the MR data, the solid red line indicates the fitting of Kohler's rule with \emph{m} = 1.8. The inset shows the field dependence of MR of SiP$_{2}$ measured near $\theta = 45^{\circ}$ at different temperatures up to 31.2 T with \emph{m} = 1.9.}\label{5}
\end{figure}
\begin{figure}
  \centering
  \includegraphics[width=8cm]{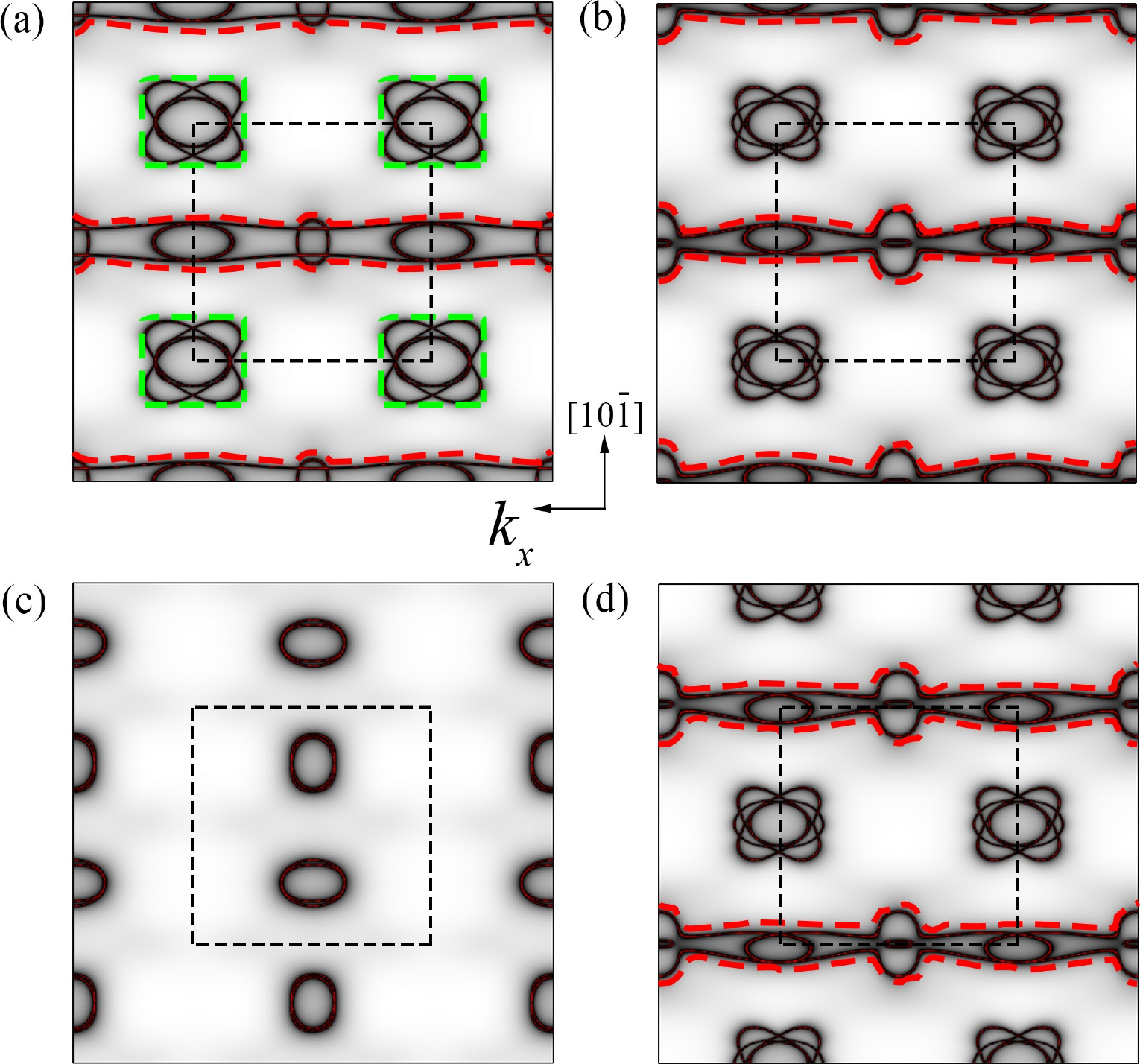}\\
  \caption{Typical cross-sections of the FS of SiP$_{2}$ projected onto the (101) plane. The horizontal axis corresponds to the $k_{x}$ direction while the vertical axis is parallel to [10$\bar{1}$]. The plane in panel (a) passes through the $\Gamma$ point, while the planes in panels (b), (c) and (d) pass through points (0, 0.1$\pi/a$, 0.1$\pi/a$), (0, 0.5$\pi/a$, 0.5$\pi/a$) and (0, 0.9$\pi/a$, 0.9$\pi/a$), respectively. The green dashed lines show closed electron orbits while the red dashed lines indicate open orbits along the $k_{x}$ direction. The black dashed squares indicate the first Brillouin zone.}\label{6}
\end{figure}

On the other hand, for this particular magnetic field orientation, incomplete compensation of the two kinds of charge carriers was confirmed by the Hall resistivity measurements. As shown in Fig. 4(a), the non-linear field dependence of Hall resistivity, $\rho_{xy}(H)$, measured at various temperatures with \emph{H} $\parallel$ \emph{a} axis, indicates its multi-bands behavior. We fitted the Hall conductivity data [see Fig. 4(b)] by using the two-band model given by \cite{ziman2001electrons}:

\begin{eqnarray}
  \sigma_{xy} = - \frac{\rho_{xy}}{\rho_{xx}^{2} + \rho_{xy}^{2}} = eB[\frac{n_{h}\mu_{h}^{2}}{1+\mu_{h}^{2}B^{2}}-\frac{n_{e}\mu_{e}^{2}}{1+\mu_{e}^{2}B^{2}}]
\end{eqnarray}

where \textit{n}$_{h}$ (\textit{n}$_{e}$) and $\mu_{h}$ ($\mu_{e}$) are the hole (electron) carrier concentrations and mobilities, respectively. The obtained \textit{n}$_{h}$ (\textit{n}$_{e}$) and $\mu_{h}$ ($\mu_{e}$) as a function of temperature are plotted in Fig. 4(c) and Fig. 4(d), respectively. It was found that \textit{n}$_{h}$ increases with decreasing temperature while \textit{n}$_{e}$ varies less with temperature. It is obvious that \textit{n}$_{h}$ is larger than \textit{n}$_{e}$ in the whole temperature range, such as \textit{n}$_{h}$ = 1.62 $\times$ 10$^{20}$ cm$^{-3}$ and \textit{n}$_{e}$ = 3.22 $\times$ 10$^{19}$ cm$^{-3}$ at 2 K, implying the incomplete compensation of both carriers. Such \textit{n}$_{h}$ (\textit{n}$_{e}$) values are similar to that in most semimetals, but higher than that in Dirac semimetals Cd$_{3}$As$_{2}$ \cite{liang2015ultrahigh}, Na$_{3}$Bi \cite{xiong2015evidence}. The $\mu_{e}$ increases with decreasing temperature at first, reaches a maximum, 1.74 $\times$ 10$^{4}$ cm$^{2}$ V$^{-1}$ s$^{-1}$, at 10 K, then drops, while the $\mu_{h}$ changes with temperature, also having a maximum near 10 K. It is important that $\mu_{e}$ is obviously larger than $\mu_{h}$ in the whole temperature range, such as $\mu_{e}$ = 1.53 $\times$ 10$^{4}$ cm$^{2}$ V$^{-1}$ s$^{-1}$, $\mu_{h}$ = 0.28 $\times$ 10$^{4}$ cm$^{2}$ V$^{-1}$ s$^{-1}$ at 2 K, shown in Fig. 4(d). In our SiP$_{2}$ crystals, the cooperative action of a substantial difference between electron and hole mobility and a moderate carrier compensation might contribute to the MR, similar to the case reported by He \emph{et al}. \cite{PhysRevLett.117.267201} for YSb, which also lacks topological protection and perfect electron-hole compensation.

According to the classical two-band model \cite{pippard1989magnetoresistance}, which does not consider the details of the topology of FSs and predicts a parabolic field dependence of MR in a compensated semimetal, a small difference of the electrons and holes densities will cause the MR to eventually saturate at higher magnetic field, such as Bi \cite{PhysRev.101.544} and graphite \cite{PhysRevB.25.5478}. In order to check the behaviors of MR at higher magnetic fields, we measured again the MR using a water-cooled magnet up to 31.2 T. Figure 2(e) presents the MR as a function of magnetic field up to 31.2 T at 1.8, 4, 8 and 10 K, the MR reaches 1.90 $\times$ 10$^{4}\%$ at 1.8 K and 31.2 T and does not show any sign of saturation up to 31.2 T. The MR also follows the Kohler scaling law described in Eq. (6) with a power exponent \emph{m} = 1.5, as shown in Fig. 2(f), rather than \emph{m} = 1.2 in the lower field region ($\mu_{0}H$ $<$ 9 T), which may result from the angular deviation ($\pm$ 7$^{\circ}$) of magnetic field orientation, due to the rotation motor limitation in our water-cooled magnet. As shown in Fig. 1(j) and the following calculation (Fig. 7), the MR is very sensitive to the magnetic field orientation as \emph{H} is applied near the \emph{a} axis, i.e., a small angular deviation near $\theta = 0^{\circ}$ [A point in Fig. 1(j)] results in a large change in MR behavior.

As \emph{H} is applied along the [101] direction [$\theta = 45^{\circ}$, the B point in Fig. 1(j)] with a maximum MR, the $\rho(T)$ measured at various fields is shown in Fig. 5(a). The resistivity is remarkably enhanced by magnetic field at lower temperatures, and the field-induced up-turn was observed, in contrast with that observed for the \emph{H} $\parallel$ \emph{a} axis, but similar to that in most materials with XMR. The normalized MR also has the same temperature dependence at various fields, as shown in Fig. 5(b). Figure 5(c) displays the MR as a function of magnetic field at various temperatures, which reaches 2.17 $\times$ 10$^{3}\%$ at 2 K and 9 T, three times larger than that for the \emph{H} $\parallel$ \emph{a} axis, and does not show any sign of saturation too. The MR can be described by the Kohler scaling law in Eq. (6) [see Fig. 5(d)] with the fitting parameters $\alpha$ = 68.4 ($\mu\Omega$ cm/T)$^{1.8}$ and \textit{m} = 1.8. The nearly quadratic field dependence of MR and the field-induced up-turn behavior are the common characteristics for most topologically nontrivial/trivial semimetals with XMR, such as WTe$_{2}$ \cite{PhysRevB.92.180402}, $\alpha$-WP$_{2}$ \cite{PhysRevB.97.245101}, $\beta$-WP$_{2}$ \cite{PhysRevB.96.121107,PhysRevLett.117.066402,PhysRevB.96.121108} and $\alpha$-Ga \cite{chen2018large} \emph{et al}., which is usually attributed to the perfect electron-hole compensation. However, it is obvious that this condition is not satisfied in our SiP$_{2}$ crystals. We check the topology of the FS projected onto the plane perpendicular to [101], as plotted in Fig. 6(a) - (d) for different planes. It is clear that the hole open orbits extending along the $k_{x}$ direction emerge. We believe that MR $\propto H^{1.8}$ for this magnetic field orientation is due to the existence of these open orbits, as discussed by Zhang \emph{et al}. \cite{PhysRevB.99.035142} for cubic Cu crystals. Also, considering the prediction of MR to saturate at higher magnetic field in the classical two-band model, as discussed above, we also measured the MR up to 31.2 T using a water-cooled magnet, as \emph{H} is applied along the [101] direction. As shown in the inset in Fig. 5(d), it was found that the MR reaches 5.88 $\times$ 10$^{4}\%$ at 1.8 K and 31.2 T, does not show any sign of saturation up to 31.2 T, too, and follows the Kohler scaling law with \emph{m} = 1.9, close to \emph{m} = 1.8 obtained from the data measured on PPMS ($<$ 9 T). It should be pointed out that the MR is not sensitive to the magnetic field orientation as \emph{H} is applied near [101] direction ($\theta = 45^{\circ}$), as shown in Fig. 1(j) and the following calculation, in contrast to the case when \emph{H} along near the \emph{a} axis ($\theta = 0^{\circ}$), although the angular deviation near $\theta = 45^{\circ}$ occurs also in the water-cooled magnet. From the above results, we conclude that the linear MR for \emph{H} $\parallel$ \emph{a} axis is attributed to incomplete carriers compensation, while the quadratic MR for \emph{H} $\parallel$ [101] results from the existence of hole open orbits.
\begin{figure}
  \centering
  \includegraphics[width=8cm]{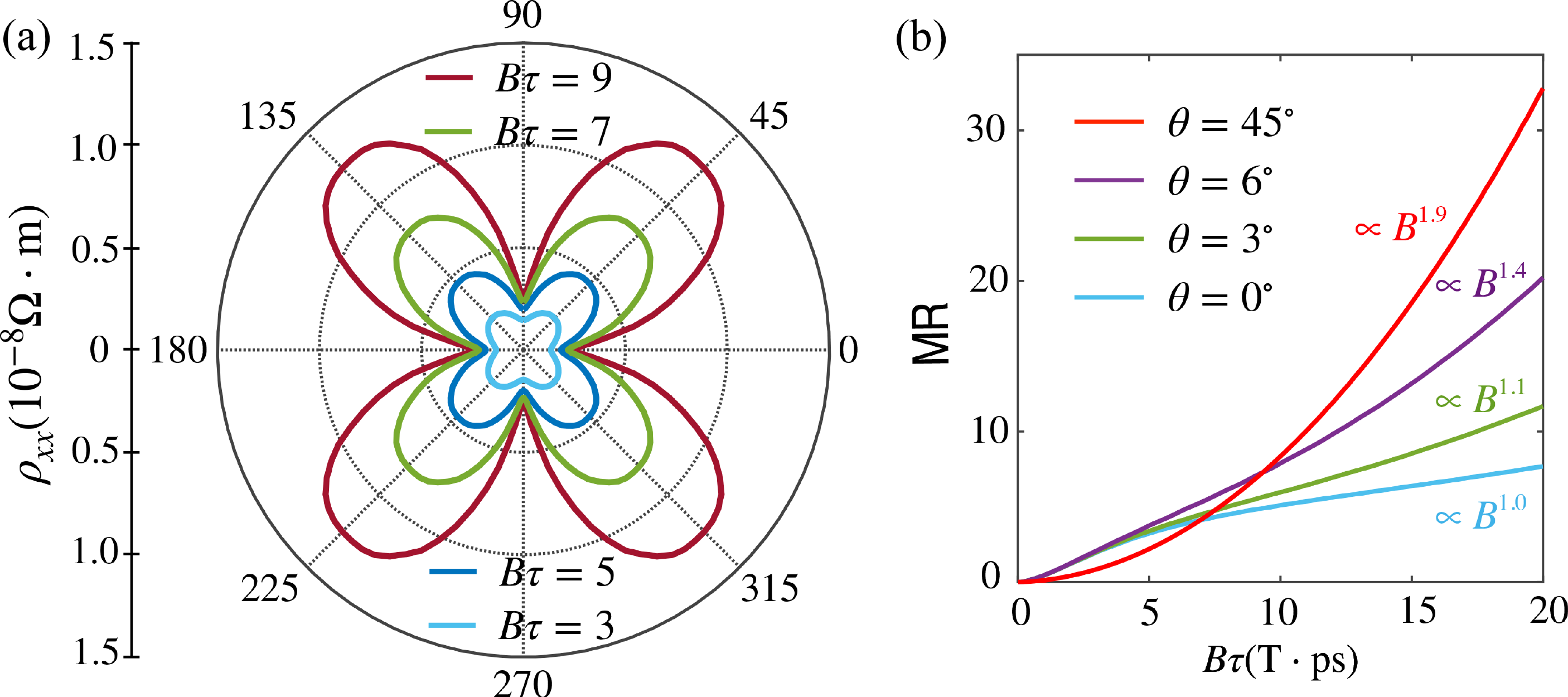}\\
  \caption{(a) Calculated anisotropy of resistivity $\rho_{xx}$ for magnetic field rotated in the $a-c$ plane agrees well with experiment results in Fig. 1(j). (b) Magnetoresistivity $\rm{MR}$ as a function of the magnitude of magnetic field for the four directions indicated by $\theta$. The resistivity at $\theta = 45^{\circ} $ is scaled by a factor of 0.25 in order to make this curve visible.}\label{7}
\end{figure}

Figure 7 shows our numerical simulation results for the resistivity anisotropy and the magnetic field dependence of MR by combining the FS discussed above with the Boltzmann transport theory approach based on the semiclassical model and the relaxation time approximation. It is clear that the calculated anisotropy of resistivity for $H$ rotated in the $a-c$ plane agrees well with the measuring results shown in Fig. 1(j). The calculated magnetic field dependence of MR also exhibits a linear behavior [see Fig. 7(b)], as $H$ oriented along the $a$ axis ($\theta$ = 0$^o$), $i.e.$, MR has $H^{1.0}$ scaling. Moreover, in case there is a misalignment of the $H$ relative to the $a$ axis, our calculations for $H$ tilting by a small angle, such as from $\theta$ = $3^o$ to $\theta$ = $6^o$, show that the magnetic field dependence of MR changes from $H^{1.1}$ to $H^{1.4}$, as shown in Fig. 7(b). All these calculated MR results for SiP$_2$ crystal, including the MR $\propto$ $H^{1.9}$ [see Fig. 7(b)] for $H$ applied in [101] direction ($\theta$ = 45$^\circ$), are well consistent with the experimental results discussed above, which indicates that the topology of FS plays the crucial role in its MR.

\begin{figure}
  \centering
  \includegraphics[width=8cm]{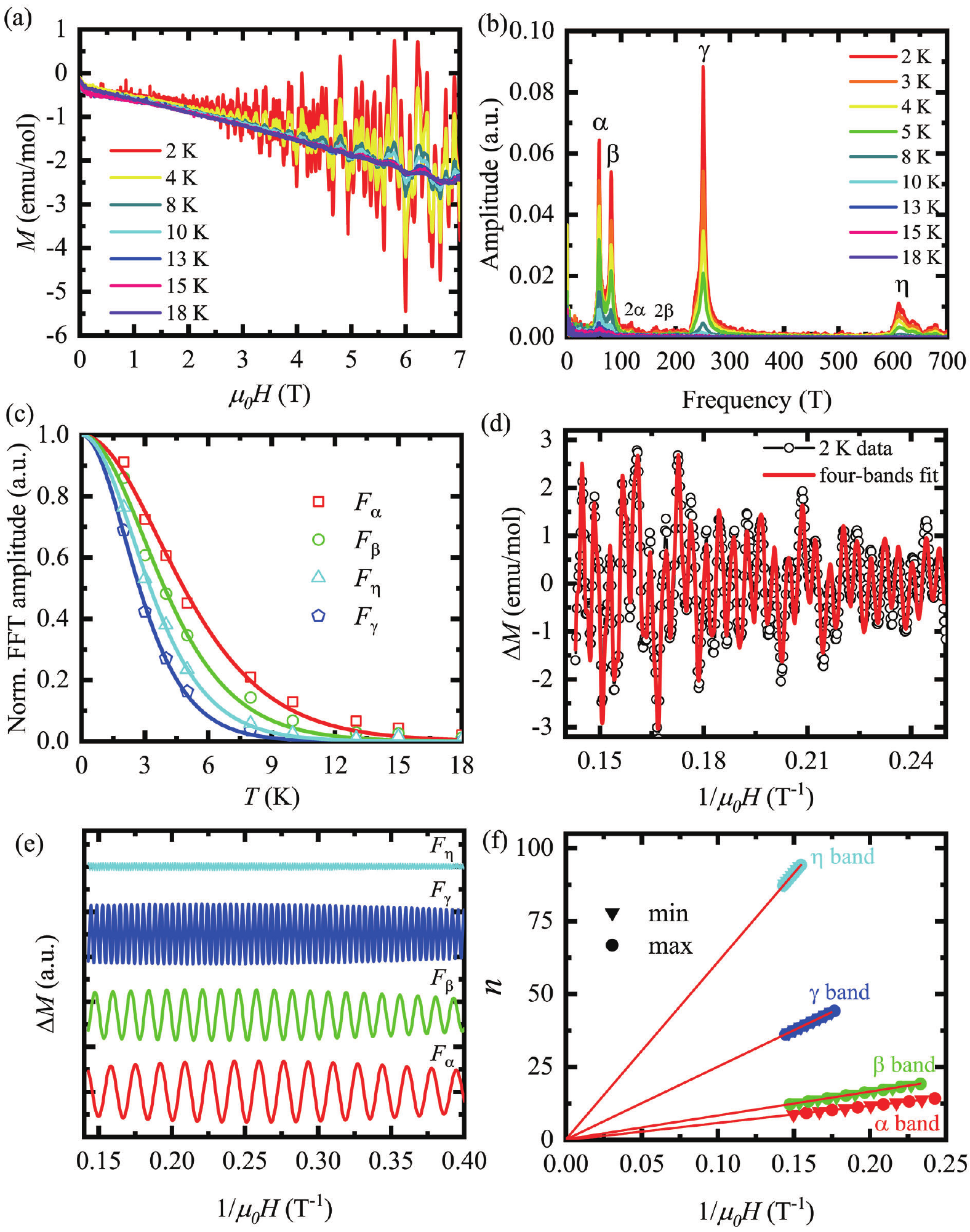}\\
  \caption{(a) The isothermal magnetization \emph{M}(\emph{H}) data with dHvA oscillations measured at various temperatures with \emph{H} applied along \textit{a} axis. (b) The FFT spectra of the oscillations at various temperatures. (c) Temperature dependence of the FFT amplitude for the four main oscillation frequencies and fitting by thermal damping relation. (d) The fitting of dHvA oscillations at 2 K by the four-bands LK formula. (e)The filtered waves of the four frequencies. (f) LL index fan diagram for the four filtered frequencies, respectively.}\label{8}
\end{figure}

Finally, in order to obtain additional information on the electronic structure, we measured the dHvA quantum oscillations in the isothermal magnetization, $M(H)$, for a  SiP$_{2}$ crystal up to 7 T for \emph{H} $\parallel$ \emph{a} axis orientation. As shown in Fig. 8(a), clear dHvA oscillations starting from 2 T in $M(H)$ curves indicate low effective masses of charge carriers. After subtracting a smooth background from the $M(H)$ data at each temperature, the periodic oscillations are visible in 1/\emph{H} up to 18 K. As an example, Figure 8(d) shows the $\Delta$\emph{M} at 2 K as a function of 1/\emph{H}. From the fast Fourier transformation (FFT) analysis, we have derived four basic frequencies \emph{F}$_{\alpha}$ (59.3 T), \emph{F}$_{\beta}$ (81.7 T), \emph{F}$_{\gamma}$ (251.1 T) and \emph{F}$_{\eta}$ (610.8 T), respectively [see Fig. 8(b)]. According to the Onsager relation \cite{shoenberg2009magnetic}: $F$ = ($\hbar/2\pi$e)$A$, we estimated the cross section area, $A$, of the FS with \emph{H} $\parallel$ \emph{a} axis, \emph{F}$_{\alpha}$ (0.00563 $\rm{\AA}^{-2}$), \emph{F}$_{\beta}$ (0.00779 $\rm{\AA}^{-2}$), \emph{F}$_{\gamma}$ (0.0239 $\rm{\AA}^{-2}$) and \emph{F}$_{\eta}$ (0.0582 $\rm{\AA}^{-2}$), respectively. The derived cross section areas of \emph{F}$_{\gamma}$ and \emph{F}$_{\eta}$ are close to the theoretical values of the hole pockets $S_{1}$ (0.0216 $\rm{\AA}^{-2}$) and $S_{2}$ (0.0519 $\rm{\AA}^{-2}$) shown in Fig. 3(a), while \emph{F}$_{\alpha}$ and \emph{F}$_{\beta}$ may originate from very small cross section area of some pockets created by a slight deviation of \emph{H} orientation from the \emph{a} axis.

In general, as discussed by Hu \emph{et al.} \cite{PhysRevB.96.045127,PhysRevLett.117.016602} for ZrSiX (X = S, Se, Te), the oscillatory magnetization for the 3D metals can be described by the Lifshitz-Kosevich (LK) formula \cite{lifshitz1956theory} with the Berry phase \cite{PhysRevLett.82.2147}:

\begin{eqnarray}
  \Delta\emph{M} \propto -B^{\frac{1}{2}}R_{T}R_{D}R_{S}\sin[2\pi(\frac{F}{B}-\gamma-\delta)]
\end{eqnarray}

where \emph{R}$_{T}$ = $\alpha$\emph{T}$\mu$/\emph{B}$\sinh$($\alpha$\emph{T}$\mu$), \emph{R}$_{D}$ = $\exp$($-\alpha$\emph{T}$_{D}$$\mu$/\emph{B}) and \emph{R}$_{S}$ = $\cos$($\pi$\textit{g}$\mu$/2), $\mu$ is the ratio of effective cyclotron mass $m^{*}$ to free electron mass $m_{0}$, the spin \textit{g}-factor \textit{g} = 2 for free electron. \emph{T}$_{D}$ is the Dingle temperature, and $\alpha$ = (2$\pi^{2}$\textit{k}$_{B}$\textit{m}$_{0}$)/($\hbar$\textit{e}). The oscillation of $\Delta$\emph{M} is described by the sine term with a phase factor $-\gamma-\delta$, in which $\gamma$ = $\frac{1}{2} - \frac{\phi_{B}}{2\pi}$ and $\phi_{B}$ is the Berry phase, the phase shift $\delta$ = $\pm$1/8 for 3D system. The effective cyclotron masses $m^{*}$ for each frequency [see Table \uppercase\expandafter{\romannumeral1}] were obtained from the fit to the temperature dependent FFT amplitudes by the thermal damping factor \emph{R}$_{T}$, as shown in Fig. 8(c). Then we used the obtained $m^{*}$ and \emph{F} values to fit the entire oscillation spectra [see Fig. 8(d)], and obtained the \emph{T}$_{D}$ and $\phi_{B}$ values (see Table \uppercase\expandafter{\romannumeral1}). For example, the \emph{T}$_{D}$ = 9.88 K for \emph{F}$_{\alpha}$, the corresponding quantum relaxation time $\tau_{Q}$ = $\hbar$/2$\pi$\textit{k}$_{B}$\emph{T}$_{D}$ = 1.23$\times$10$^{-13}$ s, the quantum mobility $\mu_{Q}$ = $e\tau_{Q}/m^{*}$ = 0.123 $\times$10$^{4}$ cm$^{2}$ V$^{-1}$ s$^{-1}$. It is important to distinguish the $\mu_{Q}$ from the transport mobility $\mu_{t}$ derived from Hall measurements. $\mu_{Q}$ is sensitive to all angle scattering processes while classical $\mu_{t}$ is only susceptible to the large angle scattering, thus $\mu_{t}$ is usually larger than $\mu_{Q}$. The Berry phase is the key feature of Dirac fermions that can be determined either directly from the multi-band fit to the LK formula or the LL fan diagram. For $\alpha$ band, the $\phi_{B}$ is estimated as 0.032$\pi$ for $\delta$ = +1/8, or $-$0.467$\pi$ for $\delta$ = $-$1/8 from the multi-band fit. Meanwhile, we filtered every single frequency from the oscillations [see Fig. 8(e)] and extracted the corresponding Berry phase from the LL index fan diagram. Generally, the integer LL indices \emph{n} should be assigned when the Fermi level lies between two adjacent LLs, where the density of state (DOS) near the Fermi level reaches a minimum, and in dHvA oscillations, the minima of $\Delta$\emph{M} should be assigned to \emph{n} $-$ 1/4 \cite{PhysRevB.86.045314,Ando}. Thus we could establish LL fan diagram as shown in Fig. 8(f). Take $\alpha$ band as an example, the extrapolation of linear fit in the LL fan diagram yields an intercept \emph{n}$_{0}$ = $-$0.1225, which corresponding to a Berry phase $\phi_{B}$ = 2$\pi$($-$0.1225 $\pm$ 1/8), and the slope of the linear fit is 59.22 corresponding to the frequency \cite{PhysRevB.96.045127,PhysRevLett.117.016602}. As shown in Table \uppercase\expandafter{\romannumeral1}, all the four bands have a similar property to the $\alpha$ band, whose Berry phase is away from $\pi$, indicating the SiP$_{2}$ is a topologically trivial semimetal.
\begin{table}
  \renewcommand\arraystretch{1.2}
  \centering
  \caption{Oscillation parameters of SiP$_{2}$}\label{1}
  \setlength{\tabcolsep}{2.5mm}
  {
  \begin{tabular}{ccccc}
    \toprule
    Parameters & \emph{F}$_{\alpha}$ & \emph{F}$_{\beta}$ & \emph{F}$_{\gamma}$ & \emph{F}$_{\eta}$\\
    \hline
    Frequency (T)& 59.3 & 81.7 & 251.1 & 610.8 \\
    \textit{m}$^{*}$/\textit{m}$_{0}$ & 0.175 & 0.216 & 0.309 & 0.268 \\
    \emph{T}$_{D}$ (K)& 9.88 & 2.42 & 2.97 & 6.74\\
    $\tau_{Q}$ (ps)& 0.123 & 0.5 & 0.407 & 0.179\\
    $\mu_{Q}$ (cm$^{2}$/Vs)& 1230 & 4074 & 2321 & 1179\\
    $\phi_{B}$ +1/8 (LK) & 0.032$\pi$ & 0.573$\pi$ & 1.072$\pi$ & 0.772$\pi$ \\
    $\phi_{B}$ $-$1/8 (LK)& $-$0.467$\pi$ & 0.073$\pi$ & 0.573$\pi$ & 0.272$\pi$   \\
    slope & 59.22&81.84&249.85&608.58\\
    intercept& -0.123&0.183&0.123&0.151\\
    $\phi_{B}$ +1/8 (LL)&0.01$\pi$&0.617$\pi$&0.495$\pi$&0.552$\pi$\\
    $\phi_{B}$ $-$1/8 (LL)&$-$0.495$\pi$&0.117$\pi$&$-$0.005$\pi$&0.052$\pi$\\
    \botrule
  \end{tabular}
  }
\end{table}

\section{\romannumeral4. CONCLUSION}
In summary, it was found that, as magnetic field is applied along the \emph{a} axis, the MR exhibits a non-saturating linear \emph{H} dependence and no field-induced up-turn behavior in resistivity emerges. The incomplete compensation of carriers was considered to be the dominant mechanism of a nearly linear \emph{H} dependence of MR. For the \emph{H} $\parallel$ [101] orientation, a non-saturating quadratic \emph{H} dependence of MR and field-induced up-turn in resistivity were observed. We argue that the existence of hole open orbits on the FS is the dominant mechanism for MR along this direction. Good agreement of the experimental results of MR with the simulations based on the FS calculated in SiP$_{2}$ indicates that the topology of FS plays the crucial role in the magnetotransport properties.

\section{ACKNOWLEDGEMENTS}

This research is supported by the Ministry of Science and Technology of China under Grants No. 2016YFA0300402 and No. 2015CB921004 and the National Natural Science Foundation of China (NSFC) (No. 11974095, 11374261), the Zhejiang Natural Science Foundation (No. LY16A040012), the Fundamental Research Funds for the Central Universities and the Chinese Academy of Science, Sharing Service Platform of CAS Large Research Infrastructure (2020-SHMFF-PT-001615). S.N.Z, Q.S.W. and O.V.Y. acknowledge the support by the NCCR Marvel. First-principles calculations were performed at the Swiss National Supercomputing Centre (CSCS) under project mr27, s832 and the facilities of Scientific IT and Application Support Center of EPFL. A portion of this work was performed on the Steady High Magnetic Field Facilities, High magnetic Field Laboratory, CAS.

\bibliography{citation}

\end{document}